\documentclass[fleqn,usenatbib]{mnras}
\usepackage{newtxtext,newtxmath}
\usepackage[T1]{fontenc}

\usepackage{graphicx}	
\usepackage{amsmath}	
\usepackage[export]{adjustbox}
\usepackage[normalem]{ulem}
\usepackage{balance}
\usepackage{flushend}
\usepackage{subcaption}
\newcommand{\pc}{\,{\rm pc}\,}
\newcommand{\kpc}{\,{\rm kpc}\,}
\newcommand{\Myr}{\,{\rm Myr}\,}
\newcommand{\Gyr}{\,{\rm Gyr}\,}
\newcommand{\kms}{\,{\rm km s}$^{-1}$\,}
\newcommand{\G}{\,{\rm G}\,}
\newcommand{\nG}{\,$10^{-9}$\G\,}
\newcommand{\g}{\,{\rm g}\,}
\newcommand{\pcc}{\,{\rm cm}$^{-3}$\,}
\newcommand{\BB}{\mathbf{B}}

\newcommand{\bb}{\mathbf{b}}

\newcommand{\uu}{\mathbf{u}}
\newcommand{\xx}{\mathbf{x}}
\newcommand{\yy}{\mathbf{y}}

\newcommand{\emf}{\mbox{\boldmath ${\cal E}$} {}}
\newcommand{\svdu}{\mbox{\boldmath ${\cal U}$} {}}
\newcommand{\svdv}{\mbox{\boldmath ${\cal V}$} {}}
\newcommand{\mean}[1]{\overline{#1}}
\newcommand{\vect}[1]{\boldsymbol{#1}}
\newcommand{\eref}[1]{Eq.~(\ref{#1})}
\newcommand{\fref}{Fig.~\,\ref}
\newcommand{\sref}{Sec.~\,\ref}
\newcommand{\aref}{Appendix~\,\ref}
\newcommand{\an}{AN}
\newcommand{\gafd}{Geophys. Astrophys. Fluid Dyn.}
\newcommand{\jfm}{J. Fluid Mech.}

\newcommand{\comment}[1]{}
\long\def\/*#1*/{}

\title[Non-local EMF]{Non-locality of the Turbulent Electromotive Force}
\author[A. B. Bendre \& K. Subramanian]{
Abhijit B. Bendre,$^{1,2}$\thanks{E-mail: abhijit.bendre@epfl.ch}
Kandaswamy Subramanian,$^{2}$\thanks{E-mail: kandu@iucaa.in}
\\
$^{1}$\'Ecole polytechnique f\'ed\'erale de Lausanne ‐ EPFL, Switzerland\\
$^{2}$IUCAA, Post Bag 4, Ganeshkhind, Pune 411007, India
}

\begin{document}
\label{firstpage}
\pagerange{\pageref{firstpage}--\pageref{lastpage}}
\maketitle

\begin{abstract}
The generation of large-scale magnetic fields
($\mean{\BB}$) in astrophysical systems is 
driven by the mean turbulent electromotive 
force ($\mean{\emf}$), the cross correlation 
between local fluctuations of velocity and 
magnetic fields. This can depend non-locally 
on $\mean{\BB}$ through a convolution kernel 
$K_{ij}$. In a new approach to find $K_{ij}$, 
we directly fit the time series data of 
$\mean{\emf}$ versus $\mean{\BB}$ from a 
galactic dynamo simulation using singular 
value decomposition. We calculate the usual 
turbulent transport coefficients as moments 
of $K_{ij}$, show the importance of including 
non-locality over eddy length-scales to fully 
capture their amplitudes and that higher order 
corrections to the standard transport 
coefficients are small in the present case. 
\end{abstract}



\section{Introduction}
A persistent theme applicable in many physical contexts 
is the influence of small-scale, or unresolved physics, 
on larger scales 
\citep{KR80,BSB16,LES_MK,GY2012,ASB17,Baumann_2012}.
This is also of crucial importance for understanding 
the origin of large-scale magnetic fields in stars and 
galaxies, ordered on scales larger than the turbulent 
motions. They are thought to be maintained by a mean 
field turbulent dynamo, through the combined action of 
helical turbulence and differential rotation. Their 
evolution is described by mean-field electrodynamics 
\citep{Radler69,Mof78,KR80,ss21,BS05}, where the 
velocity field $\mathbf{U}$ and magnetic field 
$\mathbf{B}$ are decomposed as the sums of their mean 
or large-scale (with over-line) and fluctuating or 
small-scale parts, $\mathbf{U}= \mean{\mathbf{U}}+\uu$ 
and $\mathbf{B}=\mean{\mathbf{B}} +\bb$. The mean (or 
average) is often defined over a suitable domain such 
that the Reynolds' averaging rules are satisfied. 
\footnote{These rules are: 
$\overline{\partial\BB/\partial t}=\partial\overline{
\BB}/\partial t$, $\overline{\partial\BB/\partial x_i} 
=\partial\overline{\BB}/\partial x_i$,$\overline{\BB_1 
+\BB_2}=\overline{\BB}_1+\overline{\BB}_2$,$\overline{
\overline{\BB}}=\overline{\BB}$,$\overline{\overline{B
}_i b_j}=0$ and $\overline{\overline{B}_{1i}\,
\overline{B}_{2j}}=\overline{B}_{1i}\,\overline{B}_{2j
}$.}
The evolution of $\mean{\BB}$ is then determined by the 
averaged induction equation,
\begin{align}
\frac{\partial \overline{\mathbf{B}}}{\partial t} &=\nabla \times \left(\overline{\mathbf{U}} \times \overline{\mathbf{B}} + \mean{\emf}-\eta_m \nabla \times \overline{\mathbf{B}}\right),
\label{mfe}
\end{align}
where $\eta_m$ is the microscopic diffusivity. Crucially, 
the generation of the large-scale or mean magnetic field 
(and its turbulent transport) is driven by a new contribution
in \eref{mfe}, the mean turbulent electromotive force 
(EMF), $\mean{\emf}=\mean{\uu\times\bb}$, which depends
on the cross-correlation between the turbulent velocity 
and magnetic fields. The determination of $\mean{\emf}$ 
in terms of the mean-fields themselves, either analytically 
using closure theory 
\citep{KR80,PFL76,Ditt,ss21,BF02b,RKR03,BS05} or in 
simulations, is the key to understand mean-field dynamos.

Several different approaches have been used so far to 
measure these coefficients directly from the MHD simulations 
in various contexts. \cite{cattaneoHughes}, for instance, 
estimated the coefficients $\alpha_{ij}$, for a system in 
with uniform imposed mean magnetic field which generated 
the random magnetic fields. While \cite{BranSok02} and 
\cite{kowal}, in an effort treat the additive noise in EMF 
more systematically, extracted these dynamo coefficients 
by fitting the various moments of magnetic fields (with 
EMF and themselves) with the data from direct simulations. 
Another method, developed for measuring the conductivity 
of solids, was adapted by \cite{TC13} to determine the 
large-scale diffusivity of magnetic fields, in 
two-dimensional systems.

In a more systematic approach to estimate these coefficients 
at a fixed scale, the test-field method has also been 
developed \cite[eg.][]{schriner_test,schriner_test1,axel_test}. 
This method has now been used to extract the transport 
coefficients in various contexts, such as Supernova (SN) 
driven ISM turbulence, Convective turbulence in Solar and 
geodynamo simulations, accretion disc simulations etc.
\citep[eg.][]{gressel_test,sur_test,kapala_test,bendre2015dynamo,GP15,thesis_bendre,War17}.
The test-field method relies upon a notion that the 
fluctuating fields ($\mathbf{b}_T$) generated by a set of 
imposed passive test mean magnetic fields $\mean{\mathbf{B
}}_T$ evolving with the turbulence ($\mathbf{u}$), contains 
all the information about dynamo coefficients. Components 
of the EMF associated with these test-fields ($\mean{ 
\mathbf{u}\times\mathbf{b}_T}$) are then fitted with 
test-fields (and currents) to extract all the dynamo 
coefficients, at a scale of test magnetic fields. These 
$\mathbf{b}_T$ need to be reset to their initial values 
periodically to manage the exponentially growing noise 
in the determination of the transport coefficients.

Alternatively, a straightforward approach has been also
adopted by \citet{racine,simard_svd} where they fit the 
time series of EMF to those of mean-fields and currents 
using the singular value decomposition (SVD) algorithm, 
and obtain the dynamo coefficients in the simulations 
of convection driven stellar dynamos, as the least 
square solution. Advantage of such an approach (over the 
test-field) is that it depends on the actual magnetic 
fields from the simulations rather than a fixed-scale
test-fields. A potential disadvantage is that the actual
mean-fields are more noisy than the smooth "test-fields". 
In our earlier work \citep{BK_svd_2020}, we used this 
local SVD method on a galactic dynamo simulation and in
\cite{prasun} on a thick accretion disc simulation, with
encouraging results. This motivates us to extend this 
method to also take the spatial non-locality of turbulent 
EMF into account.

This paper is structured as follows. In the following
\sref{sec:method} we introduce both local and non-local 
turbulent transport coefficients. This is followed by 
\sref{sec:simulations} which describes the setup of 
direct numerical simulations (DNS) used in this work. 
The properties of the mean and fluctuating fields 
relevant for our analysis is outlined here and in 
\aref{ISM_turb}. \sref{sec:results_kernels} describes 
the non-local SVD method for determining the transport 
coefficients. In \sref{sec:results_kernels} and 
\sref{sec:results_coefficients} we discuss the outcomes 
of this analysis, followed by conclusions in 
\sref{sec:summary}.

\section{Turbulent transport coefficients from 
\texorpdfstring{$\mean{\emf}$}{E}}
\label{sec:method}
A widely used local representation of the turbulent EMF,
assumes that $\mean{\emf}$ can be expanded in terms of 
the mean magnetic field and its derivative. In the 
current work, we use a planar, $xy$ average, to define 
mean-fields, and then the turbulent EMF can be written 
as
\begin{equation}
\mean{\emf}_{i}(z,t)=\alpha_{ij}\overline{B}_j- \eta_{ij} \left(\nabla\times\overline{\mathbf{B}}\right)_j \,,
\label{emf_local}
\end{equation}
with the indices $i$ and $j$ representing either $x$ 
or $ y$ components, and the mean-field having only a 
$z$ dependence. Here, $ \alpha_{ij}$ and $\eta_{ij}$ 
are turbulent transport tensors. When Lorentz forces 
are small and in the isotropic limit, these tensors 
are diagonal; $\alpha_{ij} = \alpha_0 \delta_{ij}$, 
$\eta_{ij}=\eta_t\delta_{ij}$. Then, in the limit of 
short correlation times, $\alpha_0$ or the $\alpha$ 
effect is proportional to the helicity of the 
turbulence and the turbulent diffusivity $\eta_t$ is 
proportional to its energy \citep{KR80,BS05,ss21}. 
Although, a number of different approaches have been 
used to estimate these coefficients even outside the 
isotropic limit, a majority assume this locality of 
the EMF. However, such an approximation is only valid 
when there is sufficient scale separation between the 
large-scale field and the turbulent velocity and 
ignores higher order derivatives of $\overline{B}_i$. 
In fact in disk galaxies, where the relevant `large' 
scale is the height of the disk of order a few 
hundred \pc while the supernovae (SN) stirring scale 
is say 100\pc, the scale separation is very modest. 
This is mirrored in the simulations of the large-scale 
dynamo in the ISM \citep{bendre2015dynamo}, where 
as we will see the scale of the mean-field is about 
200\pc while the stirring scale is of order 50-100\pc. 
Thus it is important to decipher the range of validity 
of the locality assumption.

More generally $\mean{\emf}$ can be be expressed in the 
form of a convolution with the mean-field itself 
\citep[eg.][]{Radler69,BS05,radler2014,B18},
\begin{equation}
    \mean{\emf}_i\left(z,t\right) = \int_{-\infty}^{\infty} K_{ij}\left(z,\zeta,t\right)\,\mean{B}_{j}\left(z-\zeta,t\right)\, d\zeta \,.
\label{eq_convol}
\end{equation}
This representation allows for the contribution of the 
mean magnetic field to the turbulent EMF in a non-local 
manner. For simplicity, we assume in this work that the 
time dependence is still local \footnote{See 
\cite{HB2009,RB2012} where time non-locality is explored 
using the test-field method}. Note that the convolution 
kernel $K_{ij}$ depends both on $z$ and a neighbourhood 
variable $\zeta$ separately, allowing for the 
inhomogeneity of magnetohydrodynamic turbulence in the 
general case. We will also see below that $K_{ij}$ falls 
off sufficiently rapidly with $\zeta$ that the limits of 
integration are effectively $\pm l/2$, where $l$ is order 
of the outer scale of turbulence. The more widely used 
local formulation, given by \eref{emf_local}, can be 
recovered by Taylor expanding $\overline{B}_j$ in 
\eref{eq_convol} about $z$, in powers of $\zeta$ and 
retaining the leading two terms. 
\begin{equation}
    \mean{\emf}_i = \left(\int_{-\infty}^{\infty}K_{ij}(z,\zeta)\, d\zeta\right)\, \mean{B}_j(z) - \left(\int_{-\infty}^{\infty}K_{ij}\zeta\, d\zeta \right) \,\left(\frac{\partial \mean{B}_j}{\partial z}\right)_{\zeta=0}
\end{equation}
Taking the solonoidity of mean-field into account, this 
allows the dynamo coefficients $\alpha_{ij}$ and $\eta_{ij}$ 
to be expressed in terms of the zeroth and first moments of 
$K_{ij}$ as
\begin{align}
    \begin{bmatrix} \alpha_{xx} & \alpha_{xy} \\ \alpha_{yx} & \alpha_{yy} \end{bmatrix}\ &=
    \int^{+l/2}_{-l/2} \,
    \begin{bmatrix} K_{xx} & K_{xy} \\ K_{yx} & K_{yy} \end{bmatrix}\, 
    d\zeta\nonumber\\
    \begin{bmatrix} \eta_{xx} & \eta_{xy} \\ \eta_{yx} & \eta_{yy} \end{bmatrix}\ &=
    \int^{+l/2}_{-l/2} \,
    \begin{bmatrix} -K_{xy} & K_{xx} \\ -K_{yy} & K_{yx} \end{bmatrix}\,\zeta\,d\zeta \,.
    \label{alp_eta}
\end{align}
Similarly, the subsequent higher order moments of the kernel 
multiply higher derivatives of $\overline{\BB}$. The leading 
higher order corrections to $\alpha_{ij}$ which we denote by 
$\alpha^h_{ij}$ and $\eta_{ij}$ which we denote by $\eta^h_{
ij}$ are, 
\begin{align}
    \begin{bmatrix}
    \alpha^h_{xx}&\alpha^h_{xy}\\
    \alpha^h_{yx}&\alpha^h_{yy}
    \end{bmatrix}
    =&  \int^{+l/2}_{-l/2}  
    \begin{bmatrix}
    K_{xx}&K_{xy}\\
    K_{yx}&K_{yy}
    \end{bmatrix}\zeta^2\,d\zeta
   \nonumber\\
    \begin{bmatrix}
    \eta^h_{xx}&\eta^h_{xy}\\
    \eta^h_{yx}&\eta^h_{yy}
    \end{bmatrix}
    =& \int^{+l/2}_{-l/2} 
    \begin{bmatrix}
    -K_{xy}&K_{xx}\\
    -K_{yy}&K_{yx}
    \end{bmatrix}\zeta^3\,d\zeta.
    \label{eq:higher_moments}    
\end{align}
These multiply respectively the second and third derivatives 
of $\mean{B}_j$ in the Taylor expansion of \eref{eq_convol}. 
Our aim here is to compute $K_{ij}$ from the data of direct 
numerical simulations (DNS), examine the extent of its spatial 
non-locality and also derive in a novel manner, the dynamo 
coefficients as the moments of these components. For this we 
use a galactic dynamo simulation, (performed using the NIRVANA
MHD code \cite{ZIEGLER2008227}), where supernovae (SN) 
introduced at randomly chosen points in the simulation box 
drive a multi-phase turbulent flow in the medium.

\section{Description of the direct simulations}
\begin{figure}
    \centering
    \includegraphics[valign=t,width=0.9\linewidth]{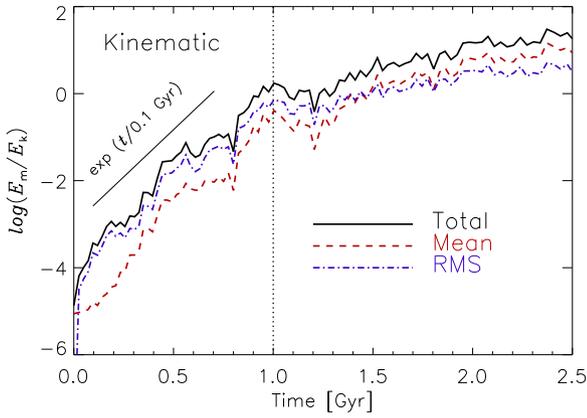}
    \caption{Time evolution of the various 
    contributions of magnetic energy relative to 
    (a nearly constant) turbulent kinetic energy 
    is plotted here. Black-solid line corresponds 
    to the total magnetic energy, red-dashed line 
    the mean magnetic energy while the blue 
    dot-dashed line shows the contribution of the 
    fluctuating part of magnetic energy relative 
    to the turbulent kinetic energy. The vertical 
    dotted line delineates the kinematic from the 
    dynamic phase of evolution and the slanted 
    solid line shows the evolution expected for 
    exponential growth of this energy ratio, for 
    a growth rate of 100\Myr.}
    \label{fig:em_t}
\end{figure}
\label{sec:simulations}
The details of the galactic dynamo simulations have been 
described in \cite{bendre2015dynamo,thesis_bendre,BK_svd_2020};
here we only summarize some features. Specifically, to have a 
reasonable comparison with the current determination of transport 
coefficients we use a run Q which has been analysed previously 
for this purpose using different methods, the test-field (TF) 
method \citep{bendre2015dynamo}, a local linear regression 
method and a local version of the singular value decomposition 
(SVD) method \citep{BK_svd_2020}. 
\begin{figure}
    \includegraphics[width=0.9\linewidth]{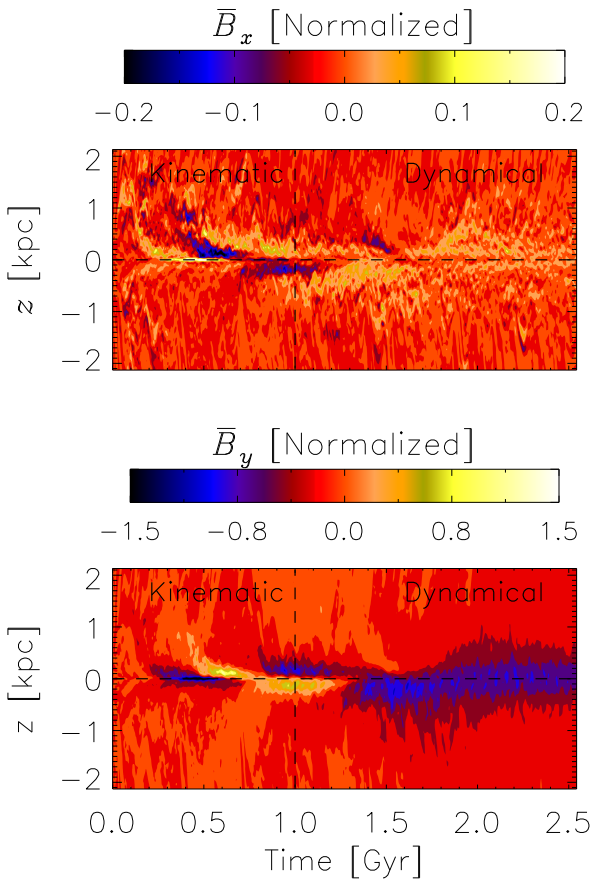}
    \caption{In the top panel we show the 
    space-time evolution of the vertical profile 
    of $\mean{{B}}_x$, along with a relevant color 
    bar on the top. In the bottom panel the same is 
    shown for the azimuthal component of $\mean{{
    B}_y}$. Color bars are normalized by the
    time-evolving square root of mean-field energy 
    to scale out the growth of magnetic field 
    components. Vertical dashed line mark roughly
    the end of initial kinematic phase, which can 
    also be seen in \fref{fig:em_t}.}
    \label{fig:b_contour}
\end{figure}
The DNS model we use is a local Cartesian shearing
box of ISM (with $L_x=L_y=0.8$\kpc, and -2.12\kpc 
to +2.12 \kpc in the $z$ direction), split in $96
\times 96\times512$ grids with a resolution of $ 
\sim 8.3$\pc. Shearing periodic boundary conditions 
are used in the radial ($x$) direction to 
incorporate the differential rotation. While the 
flat rotation curve is simulated by letting 
angular velocity scale as $\Omega\propto1/R$ 
with the radius, with a value of $\Omega_0=100$ 
\kms \kpc $^{-1}$ at the centre of the domain. In 
$z$ direction outflow conditions are used to let 
the gas escape, while preventing its inflow. 
Explicit values of kinematic viscosity ($5\times
10^{24} \rm{cm}^2 \rm{s}^{-1} $) and magnetic 
diffusivity ($2\times 10^{24} \rm{cm}^2 \rm{s}^{
-1}$) are used to avoid having the dissipation
controlled by the mesh itself, yielding a Prandtl
number of the order of 2.5. SN explosions are 
simulated as expulsions of thermal energy at 
randomly chosen locations in the box at a rate 
of $\sim 7.5$\kpc$^{-2}$\Myr$^{-1}$. Furthermore 
the distribution of SN explosion locations is also
scaled with the vertical profile of mass density. 
As an initial condition we use a vertically
stratified density profile, such that the system 
is in a hydrostatic equilibrium, with a balance
between gravitational and pressure gradient 
forces. This leads to a scale-height of $\sim300$
\pc for the density (and a midplane value of 
$10^{-24}$\g\pcc). Additionally, a piece-wise
power law is also used to describe the temperature
dependent radiative cooling. This, along with SN 
explosions, leads the plasma splitting into multiple 
thermal phases within a few \Myr, which roughly 
captures the ISM morphology. 

The initial magnetic field is of a strength of about 
\nG (about 3-4 orders of magnitude smaller than the 
equipartition strength). Both the total and mean 
magnetic field amplify exponentially with an 
e-folding time of $\sim 200$\Myr. The mean magnetic 
field goes through several reversals and parity 
changes until finally reaching to a large-scale mode 
vertically symmetric about the mid plane. Within 
about a $\Gyr$ they reach $\mu$G strengths, in 
equipartition with the turbulent kinetic energy 
density. Subsequently the magnetic energy continues 
to grow at about 5 times smaller growth rate. The 
time evolution of different magnetic energy density 
components and a space-time plot of $\overline{\BB}
(z,t)$ are shown in \fref{fig:em_t} and 
\fref{fig:b_contour}. The mean magnetic energy 
density and its space-time evolution, are also 
shown respectively in Fig. 9 and Fig. 10 of
\cite{BK_svd_2020}. An analysis of the properties 
of the fluctuating velocity and magnetic fields,
relevant for understanding the extent of
non-locality, is also given in \aref{ISM_turb}

\section{Determining the non-local kernel}
\label{sec:results_kernels}
To determine the components $K_{ij}$ using the SVD
method we proceed as follows. We first extract the
time series of $\mean{{B}}_j(z,t)$ and $\mean{\emf
}_j(z,t) =\overline{\left(\mathbf{u}\times\mathbf{ 
b } \right )}_j $ at each $  z  $ and ranging from 
approximately 0 to 900\Myr in time, corresponding 
to the kinematic phase of the dynamo (comprising 
of 800 points in total). We then express 
\eref{eq_convol} at any particular $z = z'$ in the 
form of a discrete sum,
\begin{align}
    \mean{\emf}_i (z',t) = \sum_{n= -m}^{m}\,K_{ij}(z',\zeta_n,t)\, \mean{B}_{j}(z'-\zeta_n,t) \epsilon
    \label{eq_sum_convol}
\end{align}
\begin{figure}
\centering
    \includegraphics[valign=t,width=\linewidth]{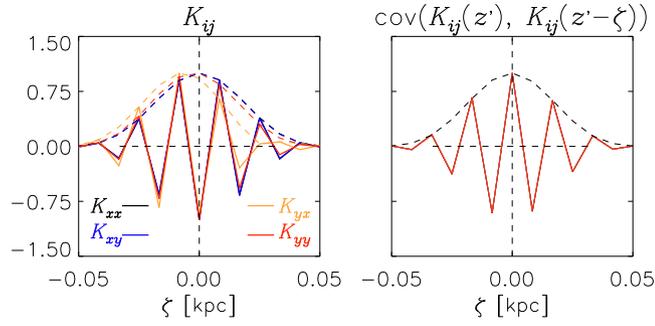}
    \caption{The left panel shows the normalized 
    values of the convolution kernel $K_{ij}$ 
    (solid lines) and its absolute value (dashed 
    lines) at a representative $z'=500$\pc. The
    right panel shows the normalized covariance 
    of $K_{xx}$ (solid line) and its absolute 
    value (dashed line) between the point 
    $\zeta=0$ (also at $z'=500$\pc) and that at 
    a neighbouring $\zeta$.}
    \label{fig:ker_cov}
\end{figure}
where $\zeta_n = n\epsilon$ denotes the location 
of mesh points in the local neighbourhood of and 
including $z'$, with the index $n$ ranging from 
$-m$ to $+m$, and $\epsilon$ the size of the
simulation mesh. The value of $m$ fixes the width
of the local neighbourhood and it is expected 
that the coefficients $K_{ij}$ would vanish for a 
large enough $ m$, such than $m \epsilon $ is 
larger than the eddy length-scales. We have 
explored vales of $m \le 6$ equivalent of a local
neighbourhood of about $-50$ to $50$\pc. This
turns out to be sufficient as discussed below and
seen from \fref{fig:ker_cov}.

For $m=6$, \eref{eq_sum_convol} then represents a
system of 2 equations in 26 unknowns (6 points on 
either sides of each $z=z'$) which we solve by 
using a time series analysis. We assume that the 
coefficients $K_{ij}$ do not vary in time during 
the initial kinematic phase of the dynamo, when
Lorentz forces are negligible which is justifiable
as discussed in our previous analysis  
\citep{BK_svd_2020}. This allows us to express 
\eref{eq_sum_convol} at each $z=z'$ and times 
$(t_1, t_2 \dots t_N)$ as an over-determined 
system of linear equations between $\mean{\emf}$ 
and $\mean{\BB}$. To solve this system for 
$K_{ij}$ using SVD, we first write 
\eref{eq_sum_convol} in a matrix form, $\mathbf{
y}_i=\mathbf{A}\cdot\mathbf{x}_i +\Hat{\mathbf{n
}}_i$ ($i$ is either $x$ or $y$), where the matrix 
$\Hat{\mathbf{n}}_i$ represents an additive noise
in the determination of the vector $\mathbf{y}_i$.
Here
\begin{align}
\centering
    \mathbf{y}_i^{\intercal} &= \begin{bmatrix} 
\mean{\emf}_i(z',t_1)&\mean{\emf}_i(z',t_2)\hdots\mean{\emf}_i(z',t_N)
\end{bmatrix}\nonumber\\
\mathbf{A}^{\intercal} 
&= \begin{bmatrix} 
\mean{B}_x(z'-6\epsilon,t_1)&
\mean{B}_x(z'-6\epsilon,t_2)&
\hdots
\mean{B}_x(z'-6\epsilon,t_N)
\\ 
\vdots&
\vdots&
\vdots\\
\mean{B}_x(z'+6\epsilon,t_1)&
\mean{B}_x(z'+6\epsilon,t_2)&
\hdots
\mean{B}_x(z'+6\epsilon,t_N)\\
\mean{B}_y(z'-6\epsilon,t_1)&
\mean{B}_y(z'-6\epsilon,t_2)&
\hdots
\mean{B}_y(z'-6\epsilon,t_N)
\\ 
\vdots&
\vdots&
\vdots\\
\mean{B}_y(z'+6\epsilon,t_1)&
\mean{B}_y(z'+6\epsilon,t_2)&
\hdots
\mean{B}_y(z'+6\epsilon,t_N)
\end{bmatrix}\nonumber\\
\mathbf{x}_i &= 
\begin{bmatrix} 
K_{ix}(z',+6\epsilon)\\
\vdots\\
K_{ix}(z',-6\epsilon)\\
K_{iy}(z',+6\epsilon)\\
\vdots\\
K_{iy}(z',-6\epsilon)
\end{bmatrix}\nonumber
\end{align}
and ${N  = 800 } $ is the length of the time 
series. Furthermore, $\mean{\BB} $ at the $m$ 
points outside the top and bottom of the 
boundaries is set to zero; however, adopting 
reflecting boundaries at the top and bottom, 
has negligible effect on the final results. 

We find the least-square solution to the matrix
relation $\mathbf{y}_i=\mathbf{A} \cdot\mathbf{
x}_i +\Hat{\mathbf{n}}_i$ by pseudo-inverting 
the design matrix $\mathbf{A} $ using the SVD 
algorithm. Specifically the matrix $\mathbf{A
}$ is represented as a singular value 
decomposition $\mathbf{A}=\svdu\,\mathbf{w}^{
-1}\, \svdv^{ \intercal}$, where $\svdu $ and 
$\svdv$ are orthonormal matrices of dimension 
($ N \times 26$), and ($ 26 \times 26$) while 
$\mathbf{ w }  $ is a $ 26\times 26$ diagonal 
matrix. The least-square solution, denoted by 
$\hat{\xx}_i$, is then determined simply by,
$\Hat{\xx}_i=\svdv\,\mathbf{w}\,\svdu^{\intercal}
\,\mathbf{y}_i$ \citep{mandel_svd,recepies}. 
Note that $\mean{\emf}$ and $\overline{\BB}$ 
grow exponentially as $\exp{(t/200{\rm Myr})}$ 
during the kinematic stage even as $K_{ij}$ 
remains constant. This growth is scaled out 
of $\yy_i$ and the columns of $\mathbf{A}$ 
before implementing the SVD algorithm to find
$\hat{\xx}_i$, as in our earlier work
\citep{BK_svd_2020}.

The SVD analysis also gives the full covariance
matrix between the $l^{th}$ and $m^{th}$ element 
of $\Hat{\xx}_i$,
\begin{align}
        \mathrm{Cov}\Big([\Hat{\xx}_i]_l,\,[\Hat{\xx}_i]_m \Big) &= \sum_{j=0}^{25} \frac{\svdv_{lj}\,\svdv_{mj}}{\mathbf{w}_{jj}^2}\,\sigma_i^2.
\label{covar}
\end{align}
Here $\sigma_{i}$ is the $1\sigma$ variance 
associated the vector $\yy_i$,
\begin{align}
\sigma_i^2 = \frac{1}{N}\left(\mathbf{y}_i - \mathbf{A}\Hat{\mathbf{x}}_i\right)^{\intercal}\left(\mathbf{y}_i - \mathbf{A}\Hat{\mathbf{x}}_i\right).
\end{align}
The diagonal ($l=m$) elements in \eref{covar} give
the variance in the $l^{th}$ element in the vector
of $\Hat{\xx}_i$ and so determines the $1\sigma$
errors in each component of the kernel $K_{ij}$. 
However, to determine the uncertainties in 
the various moments of the kernel (\eref{alp_eta}),
the fact that components $K_{ij}$ are correlated 
within the $\zeta$ neighbourhood also needs to be 
accounted for, and we use the square root of 
summation of SVD covariance matrix as a measure 
of uncertainty in the moments of kernel coefficients
determined using least-square method. For example 
consider any of the turbulent transport 
coefficients, say $Z$, which is written as the sum 
$Z=\sum_{m=-6}^{6}z_m$. Then the variance $\mathrm{
Var}(Z) = \sum_m \sum_n \mathrm{Cov}(z_n,z_m)$ and 
the measure of uncertainty in $Z$ can be calculated as $\sqrt{\mathrm{Var}(Z)}$.

\begin{figure}
    \includegraphics[width=0.95\linewidth]{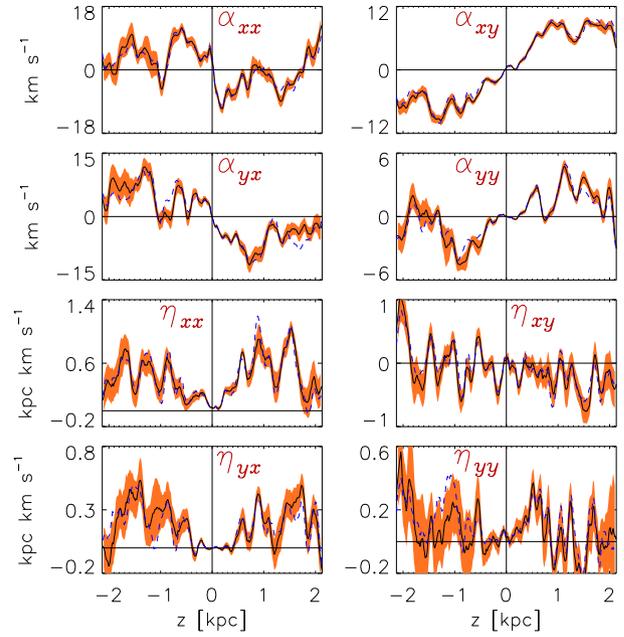}
    \caption{Blue-dashed lines show the vertical 
    profiles of the zeroth and first moments of 
    the kernel $K_{ij}$ calculated non-locally 
    using \eref{alp_eta} with the SVD method. We
    have adopted $m=6$, equivalent to the local
    neighbourhood of size $\pm50$\pc. Shown in
    black-solid lines are the same profiles but
    calculated by applying the non-local SVD
    algorithm to nine different sections of 
    time series $S_1$ to $S_9$ and averaging 
    the outcomes. Orange shaded regions 
    indicate a measure of uncertainty in these 
    coefficients obtained from these realizations 
    as described in the text.
    }
\label{fig:coeff}
\end{figure}

In the left panel of \fref{fig:ker_cov} we show 
the normalized coefficients of  the recovered 
convolution kernel $K_{ i j }  $ (solid lines) 
and their absolute values (dashed lines), as 
functions of $\zeta $ at a representative $z'=
500$\pc. We see that all coefficients drop to
negligible values within $\zeta =\pm50$\pc, 
which is also their approximate half widths. 
This scale is in fact of order of the correlation 
scale of interstellar medium turbulence in the 
simulations of \cite{bendre2015dynamo} (see the 
discussion in \aref{ISM_turb}) and also 
\cite{Hollins2017}. 

The right hand panel of \fref{fig:ker_cov} shows 
the normalised covariance (solid line) and its 
absolute value (dashed line) for $K_{xx}$, between 
its value at $\zeta=0$ (also at $z'=500$\pc) and 
that at an arbitrary neighbourhood point $\zeta$. 
Again, we see that these profiles too vanish 
smoothly at the boundaries of the chosen local 
neighbourhood. As the coefficients $K_{ij}(z',
\zeta)$ are correlated within the local 
neighbourhood, and determined only in the 
least-square sense, their amplitude at a 
particular $\zeta$ has correlated SVD variances. 
However, the extent of non-locality of
$\mean{\emf}$, is well constrained by the 
width of the dashed profiles in the left hand 
panel of \fref{fig:ker_cov}.

\begin{figure}
    \centering
    \includegraphics[width=\linewidth]{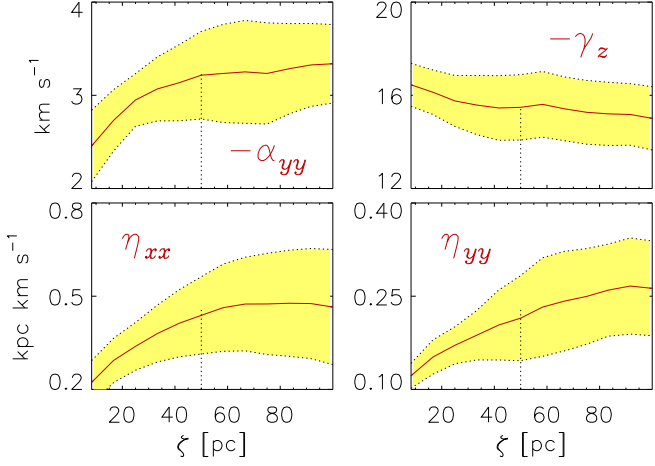}
    \caption{The solid red lines are various
    turbulent transport coefficients as functions 
    of the half width of local neighbourhood 
    (\fref{fig:coeff} corresponds to the width 
    of 50\pc). All the coefficients are averaged 
    at the vertical location $z=1$\kpc$\pm100$
    \pc, except for $\alpha_{yy}$ which is taken 
    at $z=-1$\kpc$\pm100$\pc, since its vertical 
    profile has a fluctuation at around $\sim0.8$
    \kpc, (eg. \fref{fig:coeff}). Shaded in 
    yellow are the regions corresponding to width 
    of one mean absolute deviation. Dotted 
    vertical lines at around 50\pc indicate the 
    integral scale-length of turbulence.}
    \label{fig:coeff_vs_xi}
\end{figure}

\section{turbulent transport coefficients}
\label{sec:results_coefficients}
Once the kernel components $K_{ij}(z,\zeta)$ are 
determined, the transport coefficients $\alpha_{
ij}$, $\eta_{ij}$, are obtained by calculating 
the first two moments of the kernel as in \eref{alp_eta} after converting the integrals 
to discrete sums over $\zeta_n$ with $n$ ranging 
between $\pm m$. The vertical ($z$-dependent)
profiles of these coefficients are shown in
\fref{fig:coeff} as dashed lines. The SVD
covariance in their determination, calculated 
from its variance as described above by summing 
over all the corresponding elements of the 
covariance matrix, turns out to be very small, 
less than $2-3\%$ of the coefficients themselves. 
This is both due to cancellations when summing 
over the signed covariances in $\sum_m\sum_n
\mathrm{Cov} (z_n,z_m)$, and because the term 
$\sigma_i^2$ in \eref{covar} is small as it 
depends inversely on $N$. As an alternate direct
estimate of uncertainty in these coefficients, 
we split the time series of $\mean{\mathbf{B}}$ 
and $\mean{\emf}$ into 
nine different time series ($S_1$ to $S_9$) by
extracting points that are about a correlation 
time apart (8 points in the time series) as in
\cite{BK_svd_2020}. We then use the SVD to 
determine $K_{ij}$ for each of these time 
series and compute its moments to determine
$\alpha_{ij}$ and $\eta_{ij}$. The mean of 
these 9 realizations is shown as solid lines 
in \fref{fig:coeff} and it agrees well with 
that computed from the full series. The orange
shaded regions in \fref{fig:coeff}, indicate 
the square root of variance (in these 
nine realizations) divided by the number of
realizations (see for example, \cite{recepies}). 
We note that widening the size the of local
neighbourhood amounts to the determination of 
more unknown $K_{ij}$'s in the system, which
increases uncertainties in the 
determination of transport coefficients. 
Moreover, $\overline{\emf}_y$ and $\mean{B}_x$ 
are noisier in the DNS compared to their 
counterparts, and therefore the coefficients 
$\alpha_{yx}$, $\eta_{xy}$ and $\eta_{yy}$, 
that depend on these two, are noisier compared 
to the other coefficients. 
\begin{figure}
    \centering
    \includegraphics[width=0.9\linewidth]{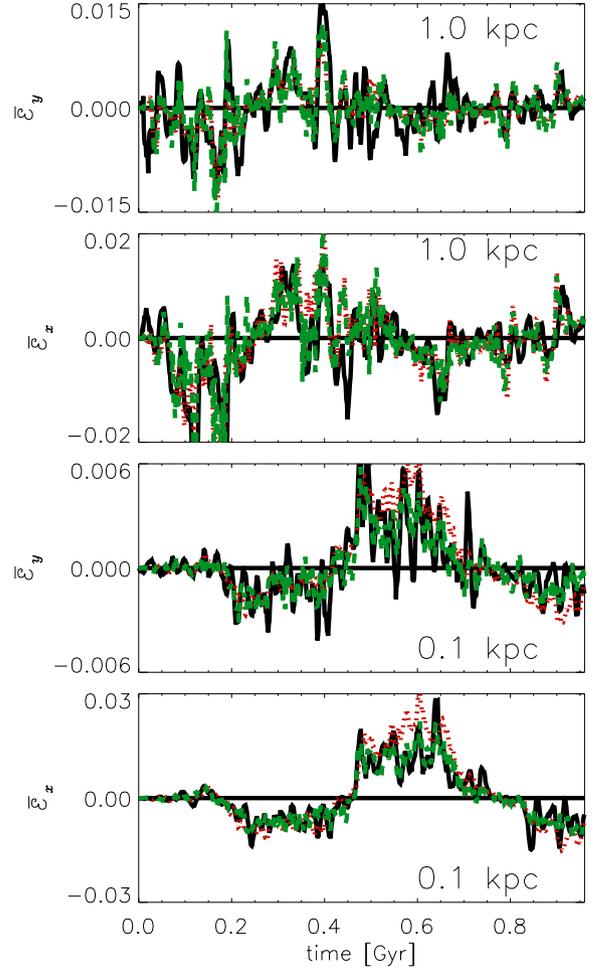}
    \caption{The black-solid lines show the time 
    series of $x$ and $y$ components of EMF, as 
    obtained from the DNS, at $z=1$\kpc (top two 
    panels), while the bottom two panels show the 
    same at $z=0.1$\kpc. The green-dashed lines 
    show the same EMF time series but as 
    reconstructed using \eref{eq_convol}, while 
    the red-dotted lines show the EMF time series
    using only first two moments of the kernel 
    (i.e. $\alpha_{ij}$ and $\eta_{ij}$). EMF
    components are expressed in the units of 
    $\mu$\G\kms, and additionally we have scaled 
    their time series with a factor $\exp{(-t/200
    {\rm Myr})}$ to account for the exponential
    growth of $\mean{\emf}_i$.}
\label{fig:emf_t_series}
\end{figure}
The $z$-dependencies of turbulent coefficients
in \fref{fig:coeff} are quite similar to their 
local determination in \cite{BK_svd_2020}; however
several of them, like $\eta_{ij}$ and 
$\alpha_{yy}$ have larger amplitudes. In fact we 
find that the vertical profiles of the dynamo 
coefficients determined adopting $m=1$ match 
very well with that determined from the same 
simulation using our previous local SVD analysis
\cite{BK_svd_2020} (see e.g. \fref{fig:coeff_non_local_m_1}). 
To examine the importance of 
non-locality more carefully, we vary the size of 
local neighbourhood in \eref{eq_sum_convol}, 
ranging from $m=1$ to 12 (about $\pm 8$\pc to 
$\pm100$\pc), and determine the components of 
$K_{ij}$ for each of those cases. The results 
are shown in \fref{fig:coeff_vs_xi} (as solid 
red lines), by averaging the coefficients at 
$z=1$\kpc$\pm100$\pc, except for $\alpha_{yy}$ 
which is averaged over $z=-1$ \kpc$\pm100$\pc. 
Shaded in yellow are the regions corresponding 
to width of one mean absolute deviation. We see 
that $\alpha_{yy}$, crucial for the generation 
of $\mean{B}_x$ from $\mean{B}_y$, and the 
turbulent diffusion coefficients $\eta_{xx}$, 
$\eta_{yy}$ all increase with the size of 
neighbourhood until $m\sim6$ (equivalent to 
$\sim \pm 50$\pc), and stabilize thereafter 
(to the profiles shown in of \fref{fig:coeff}). 
The turbulent diamagnetic pumping term 
$\gamma_z= (\alpha_{yx}-\alpha_{xy})/2$, which 
leads to a vertical advection of the mean-field 
also appears to stay constant with $m$ after 
initially decreasing up to $m\le6$. This 
indicates the importance of the non-local 
contributions included here, that are ignored 
when using \eref{emf_local} or too small a 
value of $m$ to compute the coefficients. The 
asymptotic values of these transport 
coefficients compare favorably with theoretical 
expectations for the galactic interstellar 
medium, and can lead to the large-scale dynamo 
action seen in this simulation, as already 
discussed in \cite{BK_svd_2020}. The trends 
seen in \fref{fig:coeff_vs_xi} are also 
qualitatively consistent with \cite{BRS08,RB12} 
and specifically the work of 
\cite{gressel_test_k_dep}, where the scale 
dependence of the transport coefficients 
was obtained  by varying the wave-number of
test-fields (used to measure them) from $k=1$
(equivalent to the box size of $\sim4$\kpc) to
$k=32$ (see also the discussion in 
\aref{TFM_res}). The present analysis, on the 
other hand, infers this dependence in a new 
direct approach by firstly using the actual 
mean-fields instead of test-fields, and also 
by increasing the width of the local neighbourhood, 
up from the grid size, to incorporate non-locality.
\begin{figure}
\centering
    \includegraphics[width=0.95\linewidth]{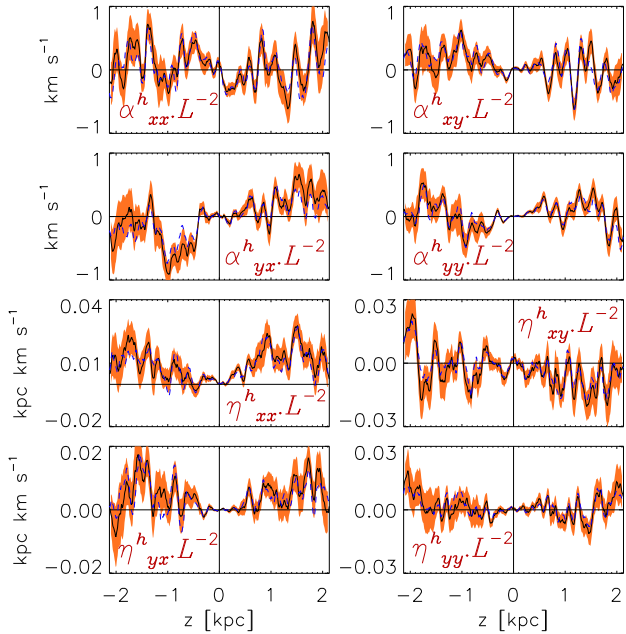}
    \caption{Shown in black-solid lines are 
    vertical profiles of $\alpha^h_{ij}$ and 
    $\eta^h_{ij}$ scaled with $L^{-2}$. These 
    are determined by applying non-local SVD 
    method to the time series $S_1$ to $S_9$, 
    same as in the \fref{fig:coeff}, and 
    averaging the outcomes. Orange regions 
    indicate the same measure of uncertainty 
    as in \fref{fig:coeff}. With blue-dashed 
    lines same profiles are shown except they 
    are calculated by taking the full time 
    series into account.}
    \label{fig:higher_moments_same_units}
\end{figure}

Finally,
in \fref{fig:emf_t_series}, we compare 
the time series of EMF components calculated 
directly from the DNS (black-solid line), with 
both the EMF reconstructed using \eref{eq_convol} 
(and the recovered kernel coefficients $K_{ij}$) 
(green-dashed line), and also that using 
\eref{emf_local} (red-dotted line) which neglects 
higher order corrections to $\alpha_{ij}$ and 
$\eta_{ij}$. This is done at two representative 
locations. From \fref{fig:emf_t_series}, it is
clear first that the non-local SVD method does 
indeed recover the EMF from the DNS reasonably 
well. Second, comparing the dashed and dotted 
lines in \fref{fig:emf_t_series}, we see that the
inclusion of higher order terms does not
significantly affect the determination of EMF, 
in the present galactic dynamo context. 
To see this explicitly, we compute $\alpha^h_{ij}$, 
and the hyper-diffusion correction $\eta^h_{ij}$ 
using \eref{eq:higher_moments} after converting it 
to discrete sums. In \fref{fig:higher_moments_same_units} 
we show these second and third moments of $K_{ij}$ 
in the same units as $\alpha_{ij}$ and $\eta_{ij}$. 
This has been done by dividing them by $L^2$, where
$L=0.2$\kpc is of order of the scale over which the
mean magnetic field varies. Note that we have not
scaled these higher order coefficients by square of 
the integral scale of turbulence, since they get 
multiplied by the higher derivatives of mean 
magnetic field in the mean-field induction equation. 
It can be seen by comparing 
\fref{fig:higher_moments_same_units} with 
\fref{fig:coeff}, that these higher order 
contributions are of order a few percent to 
ten percent of $\alpha_{ij}$ and $\eta_{ij}$. 
This can also be seen by comparing the $y$-range 
of the panels in these two figures. This is a small
correction in the present case; however such 
contributions could be important in other contexts. 

\section{Conclusions}
\label{sec:summary}
We have shown here that the turbulent EMF depends 
in a non-local manner on the mean magnetic field, 
and determined the corresponding non-local 
convolution kernel $K_{ij}(z,\zeta)$ which relates 
the two as in \eref{eq_convol}. This non-locality 
can emerge if there is only a modest scale 
separation between the scale of mean magnetic field
and the driving scale, as obtains in both disk
galaxies and in ISM simulations which realize a
large-scale dynamo. We compute these non-local 
kernels using a new approach of least-square 
fitting directly the time-series data of the 
$\mean{\emf}$ versus $\mean\BB$, from a galactic
dynamo simulation, using the SVD method. We show 
that the non-locality extends over eddy 
length-scales of order $\pm 50$\pc around any 
fiducial location and the reconstructed 
$\mean{\emf}$ using \eref{eq_convol} matches well 
with that obtained directly from the simulation. 
The lowest order moments of $K_{ij}$ over $\zeta$ 
give the standard local turbulent transport 
coefficients $\alpha_{ij}$ and $\eta_{ij}$, which
however only converge when one accounts for the 
full extent of non-locality of $K_{ij}$. Higher 
order corrections to the standard transport
coefficients are small in the present galactic 
dynamo simulation, but importantly, our method 
allows us to explicitly compute them. A caveat 
of the linear least-square fitting method using 
the SVD is that it requires $\yy_i$ and $\xx_i$ 
to vary over sufficiently large range (as in the
present case). However, their advantages are 
many; of being able to use directly the 
simulation data without having to solve a set of
auxiliary equations as in the test-field method, 
to handle additive noise $\Hat{\mathbf{n}}$, and 
to determine the full covariance matrix of the 
of the fitted parameters. As we have shown here 
the method can also be generalized to determine 
the non-locality of transport coefficients. It 
would be of interest to test this method on other
physical systems, not only in MHD and fluid 
turbulence but also in the context of any 
effective field theory, where subgrid physics 
affects larger scales.

\section*{Data Availability}
The data underlying this article will be shared 
on reasonable request to the corresponding author.

\section*{Acknowledgements} 
We thank Axel Brandenburg, Detlef Elstner, Oliver
Gressel, Aseem Paranjape, Anvar Shukurov and
particularly Jennifer Schober for very useful 
discussions and suggestions on the paper. Abhijit 
B. Bendre also thanks Jennifer Schober for 
hosting him throughout the project at EPFL. We 
thank the referee for helpful comments which has 
led to improvements in the paper.


\bibliographystyle{mnras}
\bibliography{example}

\providecommand{\noopsort}[1]{}\providecommand{\singleletter}[1]{#1}%
\begin{thebibliography}{}
\makeatletter
\relax
\def\mn@urlcharsother{\let\do\@makeother \do\$\do\&\do\#\do\^\do\_\do\%\do\~}
\def\mn@doi{\begingroup\mn@urlcharsother \@ifnextchar [ {\mn@doi@}
  {\mn@doi@[]}}
\def\mn@doi@[#1]#2{\def\@tempa{#1}\ifx\@tempa\@empty \href
  {http://dx.doi.org/#2} {doi:#2}\else \href {http://dx.doi.org/#2} {#1}\fi
  \endgroup}
\def\mn@eprint#1#2{\mn@eprint@#1:#2::\@nil}
\def\mn@eprint@arXiv#1{\href {http://arxiv.org/abs/#1} {{\tt arXiv:#1}}}
\def\mn@eprint@dblp#1{\href {http://dblp.uni-trier.de/rec/bibtex/#1.xml}
  {dblp:#1}}
\def\mn@eprint@#1:#2:#3:#4\@nil{\def\@tempa {#1}\def\@tempb {#2}\def\@tempc
  {#3}\ifx \@tempc \@empty \let \@tempc \@tempb \let \@tempb \@tempa \fi \ifx
  \@tempb \@empty \def\@tempb {arXiv}\fi \@ifundefined
  {mn@eprint@\@tempb}{\@tempb:\@tempc}{\expandafter \expandafter \csname
  mn@eprint@\@tempb\endcsname \expandafter{\@tempc}}}

\bibitem[\protect\citeauthoryear{{Aiyer}, {Subramanian}  \& {Bhat}}{{Aiyer}
  et~al.}{2017}]{ASB17}
{Aiyer} A.~K.,  {Subramanian} K.,   {Bhat} P.,  2017, \mn@doi [J.\ Fluid Mech.]
  {10.1017/jfm.2017.364}, \href
  {https://ui.adsabs.harvard.edu/abs/2017JFM...824..785A} {824, 785}

\bibitem[\protect\citeauthoryear{Baumann, Nicolis, Senatore  \&
  Zaldarriaga}{Baumann et~al.}{2012}]{Baumann_2012}
Baumann D.,  Nicolis A.,  Senatore L.,   Zaldarriaga M.,  2012, \mn@doi
  [Journal of Cosmology and Astroparticle Physics]
  {10.1088/1475-7516/2012/07/051}, 2012, 051

\bibitem[\protect\citeauthoryear{Bendre}{Bendre}{2016}]{thesis_bendre}
Bendre A.~B.,  2016, doctoralthesis, Universit{\"a}t Potsdam

\bibitem[\protect\citeauthoryear{Bendre, Gressel  \& Elstner}{Bendre
  et~al.}{2015}]{bendre2015dynamo}
Bendre A.,  Gressel O.,   Elstner D.,  2015, Astronomische Nachrichten, 336,
  991

\bibitem[\protect\citeauthoryear{{Bendre}, {Subramanian}, {Elstner}  \&
  {Gressel}}{{Bendre} et~al.}{2020}]{BK_svd_2020}
{Bendre} A.~B.,  {Subramanian} K.,  {Elstner} D.,   {Gressel} O.,  2020,
  \mn@doi [\mnras] {10.1093/mnras/stz3267}, \href
  {https://ui.adsabs.harvard.edu/abs/2020MNRAS.491.3870B} {491, 3870}

\bibitem[\protect\citeauthoryear{{Bhat}, {Subramanian}  \&
  {Brandenburg}}{{Bhat} et~al.}{2016}]{BSB16}
{Bhat} P.,  {Subramanian} K.,   {Brandenburg} A.,  2016, \mn@doi [\mnras]
  {10.1093/mnras/stw1257}, \href
  {http://adsabs.harvard.edu/abs/2016MNRAS.461..240B} {461, 240}

\bibitem[\protect\citeauthoryear{{Blackman} \& {Field}}{{Blackman} \&
  {Field}}{2002}]{BF02b}
{Blackman} E.~G.,  {Field} G.~B.,  2002, \mn@doi [\prl]
  {10.1103/PhysRevLett.89.265007}, \href
  {http://adsabs.harvard.edu/abs/2002PhRvL..89z5007B} {89, 265007}

\bibitem[\protect\citeauthoryear{{Brandenburg}}{{Brandenburg}}{2005}]{axel_test}
{Brandenburg} A.,  2005, \mn@doi [Astronomische Nachrichten]
  {10.1002/asna.200510414}, \href
  {http://adsabs.harvard.edu/abs/2005AN....326..787B} {326, 787}

\bibitem[\protect\citeauthoryear{{Brandenburg}}{{Brandenburg}}{2018}]{B18}
{Brandenburg} A.,  2018, \mn@doi [Journal of Plasma Physics]
  {10.1017/S0022377818000806}, \href
  {https://ui.adsabs.harvard.edu/abs/2018JPlPh..84d7304B} {84, 735840404}

\bibitem[\protect\citeauthoryear{{Brandenburg} \& {Sokoloff}}{{Brandenburg} \&
  {Sokoloff}}{2002}]{BranSok02}
{Brandenburg} A.,  {Sokoloff} D.,  2002, \mn@doi [Geophys.\ Astrophys.\ Fluid
  Dyn.] {10.1080/03091920290032974}, \href
  {http://adsabs.harvard.edu/abs/2002GApFD..96..319B} {96, 319}

\bibitem[\protect\citeauthoryear{{Brandenburg} \& {Subramanian}}{{Brandenburg}
  \& {Subramanian}}{2005}]{BS05}
{Brandenburg} A.,  {Subramanian} K.,  2005, \mn@doi [Physics Reports]
  {10.1016/j.physrep.2005.06.005}, \href
  {https://ui.adsabs.harvard.edu/abs/2005PhR...417....1B} {417, 1}

\bibitem[\protect\citeauthoryear{{Brandenburg}, {R{\"a}dler}  \&
  {Schrinner}}{{Brandenburg} et~al.}{2008}]{BRS08}
{Brandenburg} A.,  {R{\"a}dler} K.~H.,   {Schrinner} M.,  2008, \mn@doi [\aap]
  {10.1051/0004-6361:200809365}, \href
  {https://ui.adsabs.harvard.edu/abs/2008A&A...482..739B} {482, 739}

\bibitem[\protect\citeauthoryear{Cattaneo \& Hughes}{Cattaneo \&
  Hughes}{1996}]{cattaneoHughes}
Cattaneo F.,  Hughes D.~W.,  1996, \mn@doi [Phys. Rev. E]
  {10.1103/PhysRevE.54.R4532}, 54, R4532

\bibitem[\protect\citeauthoryear{Dhang, Bendre, Sharma  \& Subramanian}{Dhang
  et~al.}{2020}]{prasun}
Dhang P.,  Bendre A.,  Sharma P.,   Subramanian K.,  2020, \mn@doi [Monthly
  Notices of the Royal Astronomical Society] {10.1093/mnras/staa996}, 494, 4854

\bibitem[\protect\citeauthoryear{{Dittrich}, {Molchanov}, {Sokoloff}  \&
  {Ruzmaikin}}{{Dittrich} et~al.}{1984}]{Ditt}
{Dittrich} P.,  {Molchanov} S.~A.,  {Sokoloff} D.~D.,   {Ruzmaikin} A.~A.,
  1984, \an, \href {http://adsabs.harvard.edu/abs/1984AN....305Q.119D} {305,
  119}

\bibitem[\protect\citeauthoryear{Gotoh \& Yeung}{Gotoh \& Yeung}{2012}]{GY2012}
Gotoh T.,  Yeung P.,  2012, Passive Scalar Transport in Turbulence: A
  Computational Perspective.
Cambridge University Press, p. 87–131

\bibitem[\protect\citeauthoryear{{Gressel} \& {Elstner}}{{Gressel} \&
  {Elstner}}{2020}]{gressel_test_k_dep}
{Gressel} O.,  {Elstner} D.,  2020, \mn@doi [\mnras] {10.1093/mnras/staa663},
  \href {https://ui.adsabs.harvard.edu/abs/2020MNRAS.494.1180G} {494, 1180}

\bibitem[\protect\citeauthoryear{{Gressel} \& {Pessah}}{{Gressel} \&
  {Pessah}}{2015}]{GP15}
{Gressel} O.,  {Pessah} M.~E.,  2015, \mn@doi [\apj]
  {10.1088/0004-637X/810/1/59}, \href
  {https://ui.adsabs.harvard.edu/abs/2015ApJ...810...59G} {810, 59}

\bibitem[\protect\citeauthoryear{{Gressel}, {Elstner}, {Ziegler}  \&
  {R{\"u}diger}}{{Gressel} et~al.}{2008}]{gressel_test}
{Gressel} O.,  {Elstner} D.,  {Ziegler} U.,   {R{\"u}diger} G.,  2008, \mn@doi
  [Astronomy and Astrophysics] {10.1051/0004-6361:200810195}, \href
  {https://ui.adsabs.harvard.edu/abs/2008A&A...486L..35G} {486, L35}

\bibitem[\protect\citeauthoryear{{Hollins}, {Sarson}, {Shukurov}, {Fletcher}
  \& {Gent}}{{Hollins} et~al.}{2017}]{Hollins2017}
{Hollins} J.~F.,  {Sarson} G.~R.,  {Shukurov} A.,  {Fletcher} A.,   {Gent}
  F.~A.,  2017, \mn@doi [\apj] {10.3847/1538-4357/aa93e7}, \href
  {https://ui.adsabs.harvard.edu/abs/2017ApJ...850....4H} {850, 4}

\bibitem[\protect\citeauthoryear{{Hubbard} \& {Brandenburg}}{{Hubbard} \&
  {Brandenburg}}{2009}]{HB2009}
{Hubbard} A.,  {Brandenburg} A.,  2009, \mn@doi [\apj]
  {10.1088/0004-637X/706/1/712}, \href
  {https://ui.adsabs.harvard.edu/abs/2009ApJ...706..712H} {706, 712}

\bibitem[\protect\citeauthoryear{{K\"apyl\"a}, {Korpi}  \&
  {Brandenburg}}{{K\"apyl\"a} et~al.}{2009}]{kapala_test}
{K\"apyl\"a} P.~J.,  {Korpi} M.~J.,   {Brandenburg} A.,  2009, \mn@doi
  [Astronomy and Astrophysics] {10.1051/0004-6361/200811498}, 500, 633

\bibitem[\protect\citeauthoryear{{Kowal}, {Otmianowska-Mazur}  \&
  {Hanasz}}{{Kowal} et~al.}{2006}]{kowal}
{Kowal} G.,  {Otmianowska-Mazur} K.,   {Hanasz} M.,  2006, \mn@doi [\aap]
  {10.1051/0004-6361:20053582}, \href
  {https://ui.adsabs.harvard.edu/abs/2006A&A...445..915K} {445, 915}

\bibitem[\protect\citeauthoryear{{Krause} \& {R{\"a}dler}}{{Krause} \&
  {R{\"a}dler}}{1980}]{KR80}
{Krause} F.,  {R{\"a}dler} K.-H.,  1980, {Mean-Field Magnetohydrodynamics and
  Dynamo Theory}.
Pergamon Press (also Akademie-Verlag: Berlin), Oxford

\bibitem[\protect\citeauthoryear{Mandel}{Mandel}{1982}]{mandel_svd}
Mandel J.,  1982, \mn@doi [The American Statistician]
  {10.1080/00031305.1982.10482771}, 36, 15

\bibitem[\protect\citeauthoryear{Meneveau \& Katz}{Meneveau \&
  Katz}{2000}]{LES_MK}
Meneveau C.,  Katz J.,  2000, \mn@doi [Annual Review of Fluid Mechanics]
  {10.1146/annurev.fluid.32.1.1}, 32, 1

\bibitem[\protect\citeauthoryear{{Moffatt}}{{Moffatt}}{1978}]{Mof78}
{Moffatt} H.~K.,  1978, {Magnetic Field Generation in Electrically Conducting
  Fluids}.
Cambridge Univ.\ Press, Cambridge

\bibitem[\protect\citeauthoryear{{Pouquet}, {Frisch}  \& {Leorat}}{{Pouquet}
  et~al.}{1976}]{PFL76}
{Pouquet} A.,  {Frisch} U.,   {Leorat} J.,  1976, \mn@doi [J.\ Fluid Mech]
  {10.1017/S0022112076002140}, \href
  {http://adsabs.harvard.edu/abs/1976JFM....77..321P} {77, 321}

\bibitem[\protect\citeauthoryear{Press, Teukolsky, Vetterling  \&
  Flannery}{Press et~al.}{1992}]{recepies}
Press W.~H.,  Teukolsky S.~A.,  Vetterling W.~T.,   Flannery B.~P.,  1992,
  Numerical Recipes in C (2Nd Ed.): The Art of Scientific Computing.
Cambridge University Press, New York, NY, USA

\bibitem[\protect\citeauthoryear{{Racine}, {Charbonneau}, {Ghizaru}, {Bouchat}
  \& {Smolarkiewicz}}{{Racine} et~al.}{2011}]{racine}
{Racine} {\'E}.,  {Charbonneau} P.,  {Ghizaru} M.,  {Bouchat} A.,
  {Smolarkiewicz} P.~K.,  2011, \mn@doi [\apj] {10.1088/0004-637X/735/1/46},
  \href {https://ui.adsabs.harvard.edu/abs/2011ApJ...735...46R} {735, 46}

\bibitem[\protect\citeauthoryear{{R{\"a}dler}}{{R{\"a}dler}}{1969}]{Radler69}
{R{\"a}dler} K.-H.,  1969, Veroeffentlichungen der Geod.~Geophys, \href
  {http://adsabs.harvard.edu/abs/1969VeGG...13..131R} {13, 131}

\bibitem[\protect\citeauthoryear{{R{\"a}dler}}{{R{\"a}dler}}{2014}]{radler2014}
{R{\"a}dler} K.~H.,  2014, arXiv e-prints, \href
  {https://ui.adsabs.harvard.edu/abs/2014arXiv1402.6557R} {p. arXiv:1402.6557}

\bibitem[\protect\citeauthoryear{{R{\"a}dler}, {Kleeorin}  \&
  {Rogachevskii}}{{R{\"a}dler} et~al.}{2003}]{RKR03}
{R{\"a}dler} K.-H.,  {Kleeorin} N.,   {Rogachevskii} I.,  2003, \mn@doi [\gafd]
  {10.1080/0309192031000151212}, \href
  {http://adsabs.harvard.edu/abs/2003GApFD..97..249R} {97, 249}

\bibitem[\protect\citeauthoryear{{Rheinhardt} \& {Brandenburg}}{{Rheinhardt} \&
  {Brandenburg}}{2012a}]{RB2012}
{Rheinhardt} M.,  {Brandenburg} A.,  2012a, \mn@doi [Astronomische Nachrichten]
  {10.1002/asna.201111625}, \href
  {https://ui.adsabs.harvard.edu/abs/2012AN....333...71R} {333, 71}

\bibitem[\protect\citeauthoryear{{Rheinhardt} \& {Brandenburg}}{{Rheinhardt} \&
  {Brandenburg}}{2012b}]{RB12}
{Rheinhardt} M.,  {Brandenburg} A.,  2012b, \mn@doi [Astronomische Nachrichten]
  {10.1002/asna.201111625}, \href
  {https://ui.adsabs.harvard.edu/abs/2012AN....333...71R} {333, 71}

\bibitem[\protect\citeauthoryear{{Schrinner}, {R{\"a}dler}, {Schmitt},
  {Rheinhardt}  \& {Christensen}}{{Schrinner} et~al.}{2005}]{schriner_test}
{Schrinner} M.,  {R{\"a}dler} K.-H.,  {Schmitt} D.,  {Rheinhardt} M.,
  {Christensen} U.,  2005, \mn@doi [Astronomische Nachrichten]
  {10.1002/asna.200410384}, \href
  {http://adsabs.harvard.edu/abs/2005AN....326..245S} {326, 245}

\bibitem[\protect\citeauthoryear{{Schrinner}, {R{\"a}dler}, {Schmitt},
  {Rheinhardt}  \& {Christensen}}{{Schrinner} et~al.}{2007}]{schriner_test1}
{Schrinner} M.,  {R{\"a}dler} K.-H.,  {Schmitt} D.,  {Rheinhardt} M.,
  {Christensen} U.~R.,  2007, \mn@doi [Geophysical and Astrophysical Fluid
  Dynamics] {10.1080/03091920701345707}, \href
  {http://adsabs.harvard.edu/abs/2007GApFD.101...81S} {101, 81}

\bibitem[\protect\citeauthoryear{{Shukurov} \& {Subramanian}}{{Shukurov} \&
  {Subramanian}}{2021}]{ss21}
{Shukurov} A.,  {Subramanian} K.,  2021, Astrophysical Magnetic Fields: From
  Galaxies to the Early Universe.
Cambridge Univ.\ Press, Cambridge

\bibitem[\protect\citeauthoryear{{Simard}, {Charbonneau}  \&
  {Dub{\'e}}}{{Simard} et~al.}{2016}]{simard_svd}
{Simard} C.,  {Charbonneau} P.,   {Dub{\'e}} C.,  2016, \mn@doi [Advances in
  Space Research] {10.1016/j.asr.2016.03.041}, \href
  {https://ui.adsabs.harvard.edu/abs/2016AdSpR..58.1522S} {58, 1522}

\bibitem[\protect\citeauthoryear{{Sur}, {Brandenburg}  \& {Subramanian}}{{Sur}
  et~al.}{2008}]{sur_test}
{Sur} S.,  {Brandenburg} A.,   {Subramanian} K.,  2008, \mn@doi [\mnras]
  {10.1111/j.1745-3933.2008.00423.x}, \href
  {http://adsabs.harvard.edu/abs/2008MNRAS.385L..15S} {385, L15}

\bibitem[\protect\citeauthoryear{{Tobias} \& {Cattaneo}}{{Tobias} \&
  {Cattaneo}}{2013}]{TC13}
{Tobias} S.~M.,  {Cattaneo} F.,  2013, \mn@doi [\jfm] {10.1017/jfm.2012.575},
  \href {https://ui.adsabs.harvard.edu/abs/2013JFM...717..347T} {717, 347}

\bibitem[\protect\citeauthoryear{{Warnecke}, {Rheinhardt}, {Tuomisto},
  {K{\"a}pyl{\"a}}, {K{\"a}pyl{\"a}}  \& {Brandenburg}}{{Warnecke}
  et~al.}{2018}]{War17}
{Warnecke} J.,  {Rheinhardt} M.,  {Tuomisto} S.,  {K{\"a}pyl{\"a}} P.~J.,
  {K{\"a}pyl{\"a}} M.~J.,   {Brandenburg} A.,  2018, \mn@doi [\aap]
  {10.1051/0004-6361/201628136}, \href
  {https://ui.adsabs.harvard.edu/abs/2018A&A...609A..51W} {609, A51}

\bibitem[\protect\citeauthoryear{Ziegler}{Ziegler}{2008}]{ZIEGLER2008227}
Ziegler U.,  2008, \mn@doi [Computer Physics Communications]
  {https://doi.org/10.1016/j.cpc.2008.02.017}, 179, 227

\makeatother
\end{thebibliography}


\appendix

\section{Properties of the ISM turbulence}
\label{ISM_turb}
\begin{figure}
    \centering
    \includegraphics[width=0.9\linewidth]{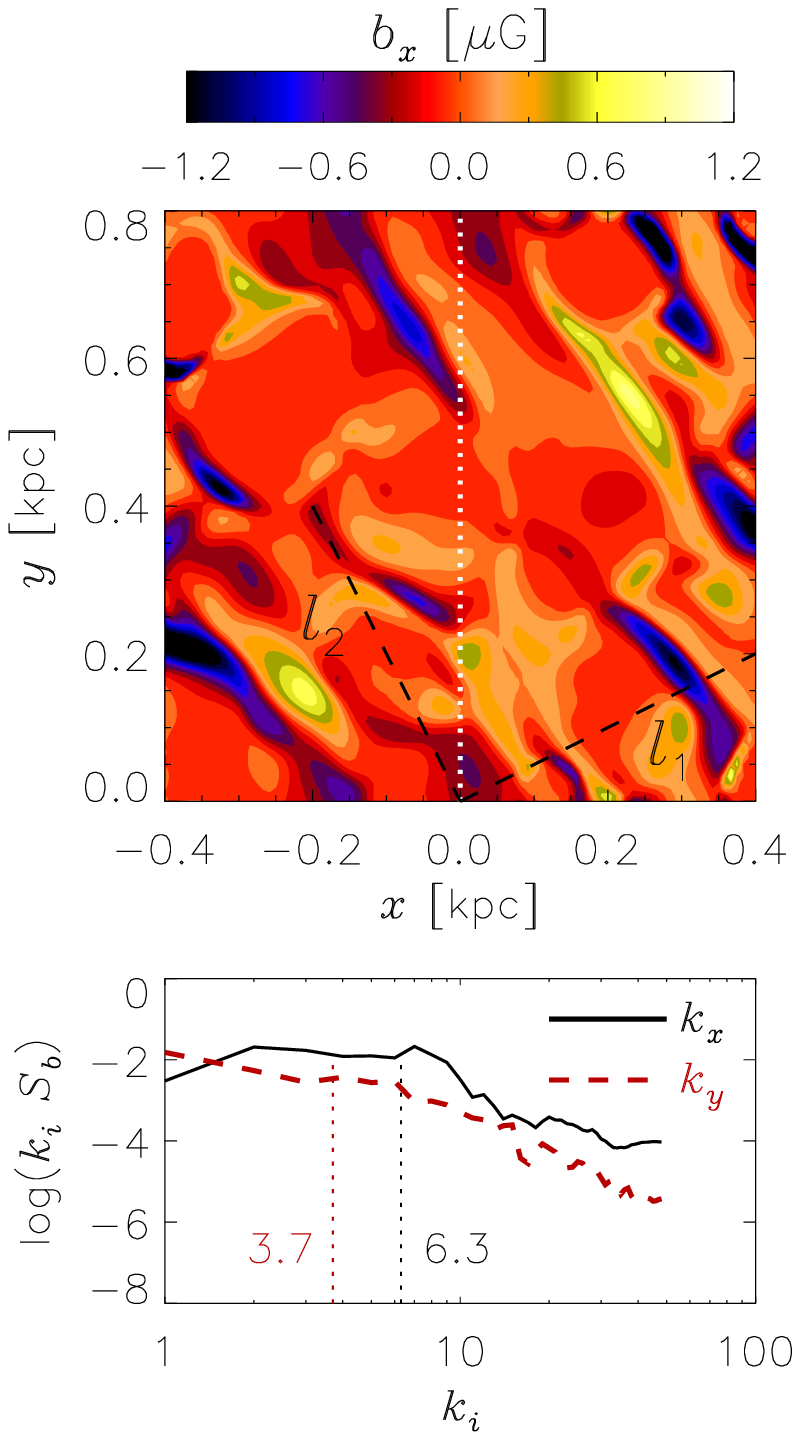}
    \caption{In the top panel we show contours of the 
    $x$ component of fluctuating magnetic field with a 
    corresponding color bar. While, in the bottom panel 
    we plot the power per unit logarithmic interval in 
    $k$-space of $\mathbf{b}$ as functions of $k_x$ with
    $k_y=1$ (black-solid line) and also as a function of
    $k_y$ with $k_x=1$ (red-dashed line). Vertical 
    (dotted) lines below the curves indicate the average
    wavenumber of the spectra. The calculation is 
    performed in 2 different directions since the field
    itself is anisotropic as can also be inferred from 
    the contours.}
    \label{fig:b_spctm}
\end{figure}
\begin{figure}
    \centering
    \includegraphics[width=0.9\linewidth]{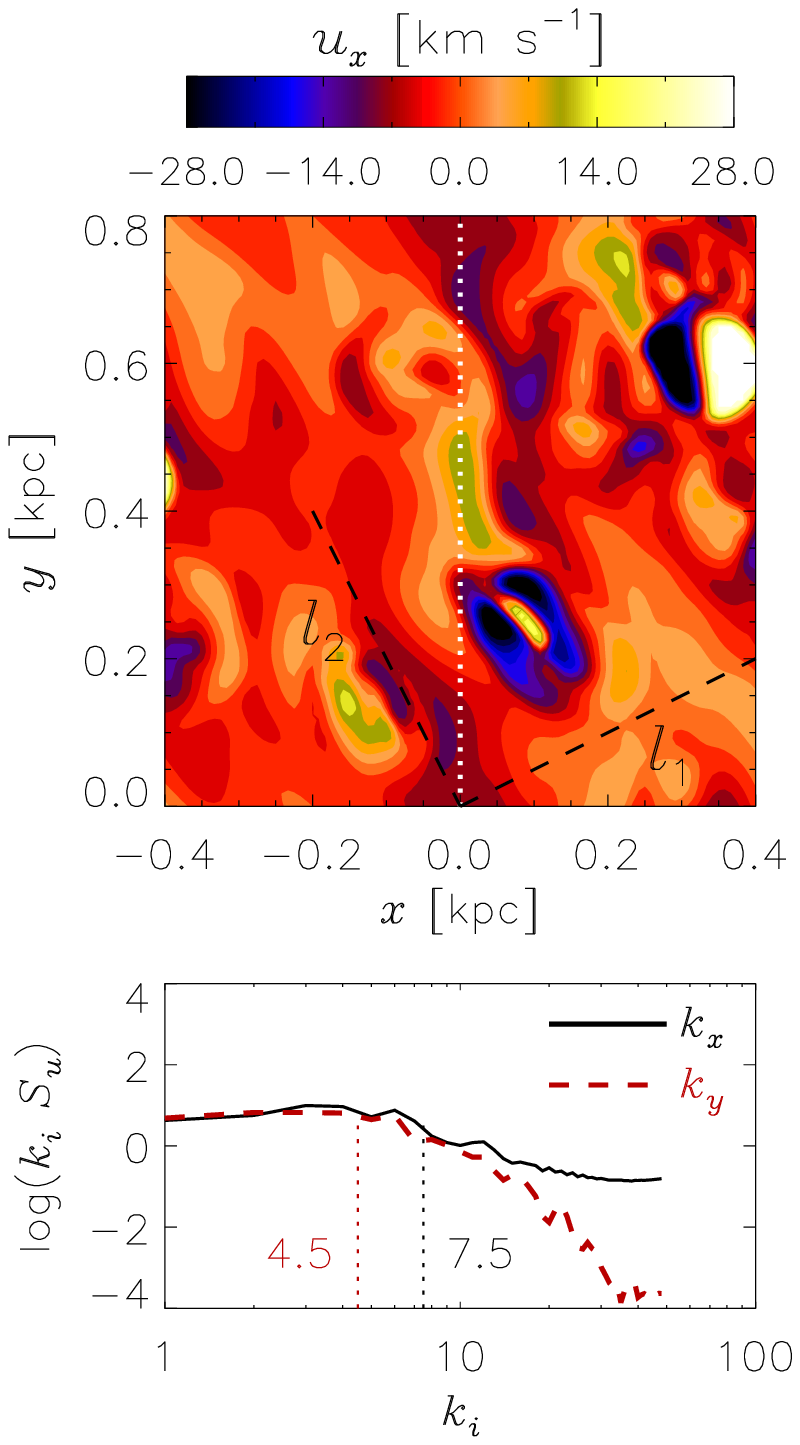}
    \caption{same as \fref{fig:b_spctm} but for $\mathbf{u}$}
    \label{fig:u_spctm}
\end{figure}

It is of interest to determine the spectral and 
correlation properties of the turbulent velocity 
and magnetic fields which lead to the mean 
electromotive force. This will also aid in seeing 
the relevance of 50\pc scale appearing in the 
half widths of $K_{ij}(\zeta)$ and its connection 
to length-scales of the turbulence directly. We 
first analyze the two dimensional power spectra 
of $\mathbf{b}$ (and also for $\mathbf{u}$) over 
$xy$ planes at several different heights $z$. We 
define the power spectrum of a quantity 
$q(\vect{r})$ as,
\begin{align}
    S_q(k_x,k_y) = 
    \left|\,\int_{\vect{r}} q(\vect{r})\exp{\left(-2\pi i \vect{k}\cdot\vect{r}\right)}\, d^2\vect{r}\,\right|^2
    \label{eq:pw_spct}
\end{align}
Where $\vect{r}$ is a vector in $xy$ plane, 
$\vect{k}$ a vector in $(k_x,k_y)$ plane and 
$q(\vect{r})$ can be any component of 
$\mathbf{b}(x,y) $ or $\mathbf{u}(x,y)$. The 
usual practice is then to define shell averaged 
1-D spectra. However, the turbulence itself 
is anisotropic in the present case, due to 
the presence of differential shear. This may 
also be realized qualitatively just by noting 
the presence of elongated structures in the 
contours of both $b_x(x,y)$ and $u_x(x,y)$ 
shown in the top panels of \fref{fig:b_spctm} 
and \fref{fig:u_spctm} with two distinct 
integral-scales roughly along the direction 
shown by $l_1$ and $l_2$. Consequently, the
power-spectra of $\mathbf{b}$ and $\mathbf{u}$ 
($S_b$ and $S_u$, defined as the sum of the 
power spectra of their components) along these 
two directions yield different integral 
wave-numbers (larger along the direction of 
$l_1$). We show sections of the anisotropic
2-dimensional power spectrum of $\mathbf{b}$ 
and $\mathbf{u}$, in the bottom panels of 
\fref{fig:b_spctm} and \fref{fig:u_spctm} 
respectively. Specifically the solid lines 
show the power per unit logarithmic interval 
in $k$-space along $k_x$ with $k_y=1$ and 
the dashed lines along $k_y$ with $k_x=1$. 
This anisotropy is primarily the reason why 
refrain from computing the $k$-shell averaged 
power spectra over $(k_x,k_y)$ plane or the 
integral-scales using these power-spectra.

\begin{figure*}
    \centering
    \includegraphics[width=0.9\textwidth]{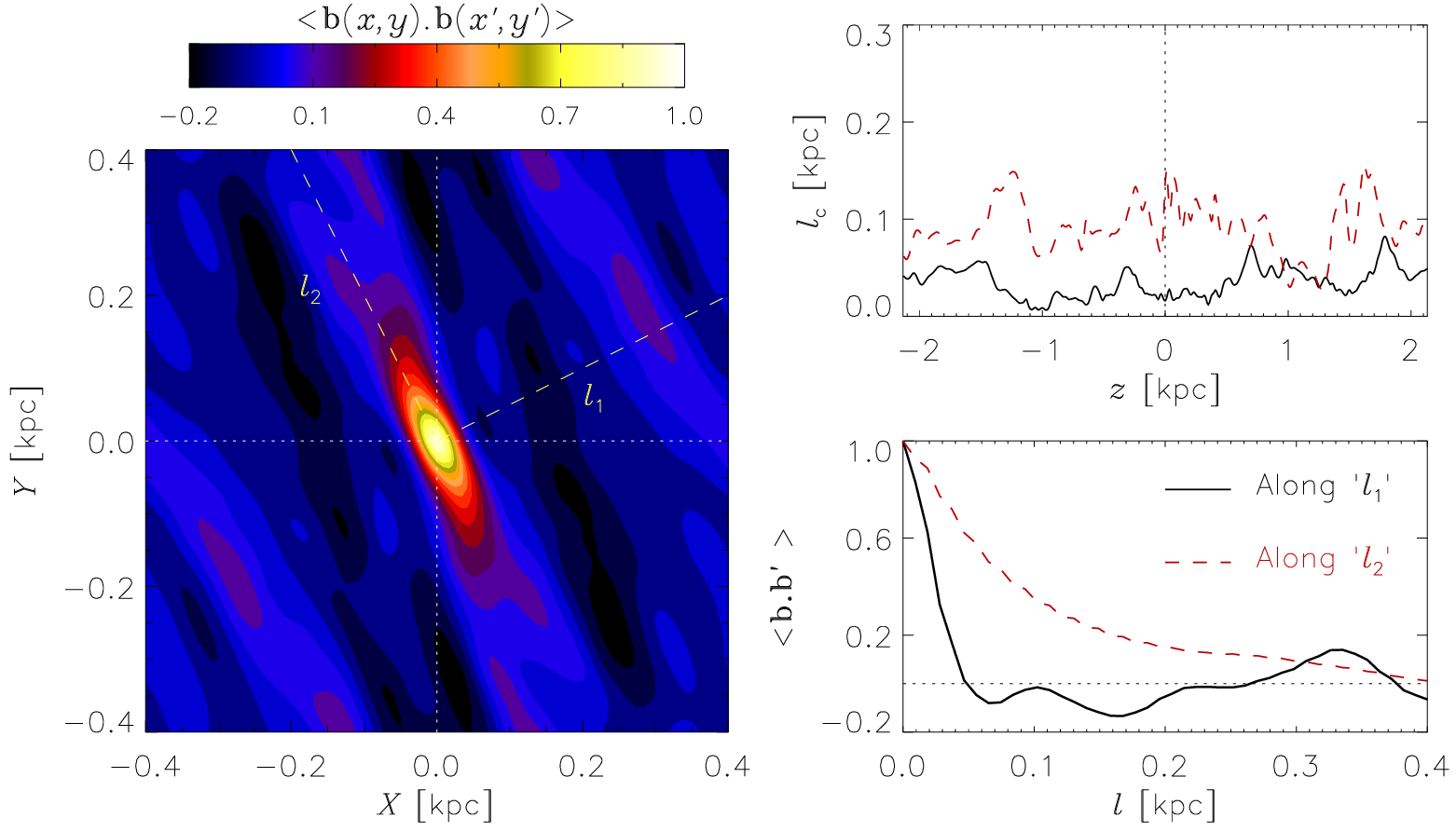}
    \caption{
    In the left hand panel we show the contour of the
    normalized 2-point correlation function $ C_b(X,Y) 
    /C_b(0,0)$, at $z=-0.8$\kpc, along with a 
    corresponding color bar. Elliptical shape of the 
    contour levels revels the statistical anisotropy 
    of $\mathbf{b}$ due to shear. Integration of this
    function along a particular direction provides a 
    correlation length-scale, $l_c$, along it. In the 
    right hand lower panel we plot these normalized
    correlations along the direction $l_1$  (minor 
    axis of the ellipse) with a black-solid line. 
    While the red-dashed line indicates the same along 
    the direction of $l_2$. In the upper half of 
    right-hand panel these correlation lengths, along 
    $l_1$ and $l_2$ are shown as the function of $z$, 
    with a same color code.}
    \label{fig:b_corr}
\end{figure*}

\begin{figure*}
    \centering
    \includegraphics[width=0.9\textwidth,keepaspectratio]{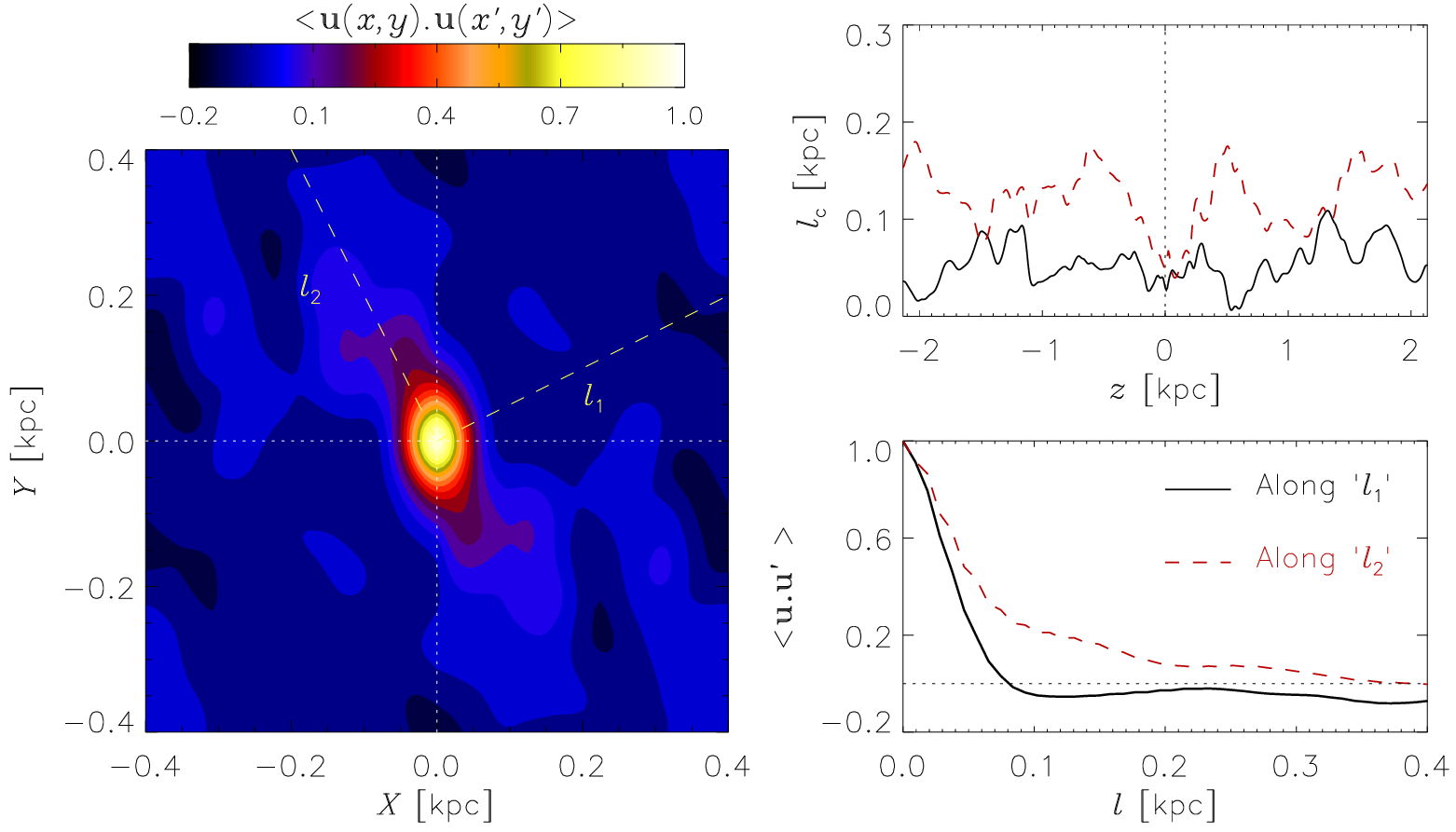}
    \caption{Same as \fref{fig:b_corr} but for $\mathbf{u}$.}
    \label{fig:u_corr}
\end{figure*}

In order to compute the integral-scales of 
$\mathbf{b}$ and $\mathbf{u}$ along $l_1$ 
and $l_2$, we adopt an alternate approach. 
We calculate the two point correlation
functions for $\mathbf{b}$ and $\mathbf{u}$ 
in $xy$ planes, $C_b$ and $C_u$, at various 
heights $z$, integral of which along any 
specific direction reflects the average 
correlation length-scale in that direction. 
The two point correlation function of 
$q(\vect{r})$ is given by the Fourier 
transform of the corresponding power-spectrum
\begin{equation}
  C_q(\vect{R}) =\,<q(\vect{r})\cdot q(\vect{r}')>
\,= \int_{\vect{k}}\, S_q(\vect{k})
 \exp{\left(2\pi i \vect{k}\cdot\vect{R}\right)}\,d^2\vect{k}\, ,
 \label{corrq}
\end{equation}
where $\vect{R}=\vect{r} - \vect{r}'$ is the 
relative position vector with components 
$(X,Y)$. Specifically, we define the 
correlation function of $\mathbf{b}$ to be
$C_b(X,Y) = <\mathbf{b}(\vect{r})\cdot
\mathbf{b}(\vect{r}')>$ calculated by replacing 
$S_q$  in \eref{corrq} by $S_b$, and 
correspondingly define $ C_u(X,Y) $ for the 
velocity field $\vect{u}$. In the left hand 
panels of \fref{fig:b_corr} and 
\fref{fig:u_corr}, we show as a contour plot, 
these correlations functions (normalized with 
respect to the value at origin), along with 
respective color bars. It can be seen that 
averaged correlation length (or the line 
integral of normalized correlation functions) 
along the direction $l_1$ and orthogonal 
direction $l_2$ are indeed different. In the
bottom panel on the right hand side of each 
figure, we plot these correlation functions 
against length along $l_1$ (black-solid lines) 
and $l_2$ (red-dashed lines). We integrate the 
normalized $C_b$ and $C_u$ along $l_1$ and 
$l_2$ to compute the averaged correlation 
lengths in those directions and repeat this 
analysis at different heights ($z$). The top 
panel on the right-hand side (of 
\fref{fig:b_corr} and \fref{fig:u_corr}) 
shows these averaged correlation lengths 
along $l_1$ (in black-solid lines) and $l_2$ 
in (red-dashed lines) as functions of $z$. 
They are $\sim 50$\pc and $\sim 100$\pc 
respectively, for both $\mathbf{b}$ and 
$\mathbf{u}$. This also yields magnetic 
Reynolds number $R_m$ of the order of 70-150 
(the magnetic Prandtl number is $\sim2.5$).
A similar analysis is also performed to 
compute the correlation length-scales in the 
$z$ directions however assuming homogeneity 
in $z$ direction, despite the vertical 
stratification and they are $\sim50$\pc, 
similar to that along $l_1$. These correlation 
lengths are consistent with the half widths of 
$K_{ij}(\zeta)$ inferred in 
\sref{sec:results_kernels} and the extent of 
non-locality in the transport coefficients 
found in \sref{sec:results_coefficients}. 

\section{Comparison with Test-Field and local SVD results}
\label{TFM_res}
\begin{figure*}
\centering
\begin{subfigure}[t]{0.45\textwidth}
    \centering
    \includegraphics[width=1\linewidth]{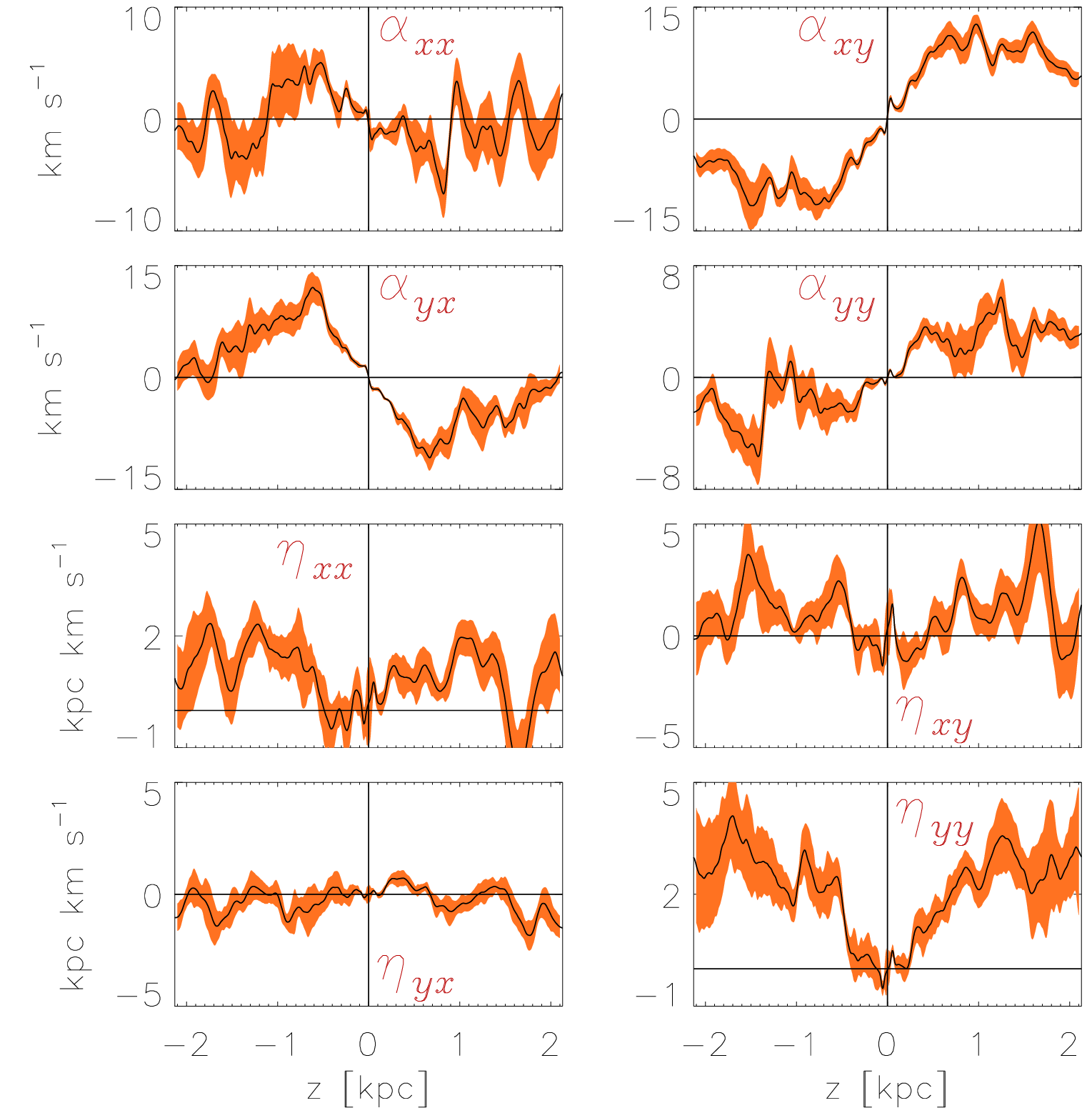}
    \caption{Shown with black-solid lines are 
    various dynamo coefficients for the same model
    as used in the main text, obtained using the 
    test-field method, by dividing the whole 
    time-series in nine sections and taking the 
    averages of the outcomes. Shaded in orange 
    are regions corresponding to uncertainties 
    in these determinations obtained the same 
    way as in \fref{fig:coeff}}
    \label{fig:test-field}
\end{subfigure}\hfill
\begin{subfigure}[t]{0.45\textwidth}
    \centering
    \includegraphics[width=1\linewidth]{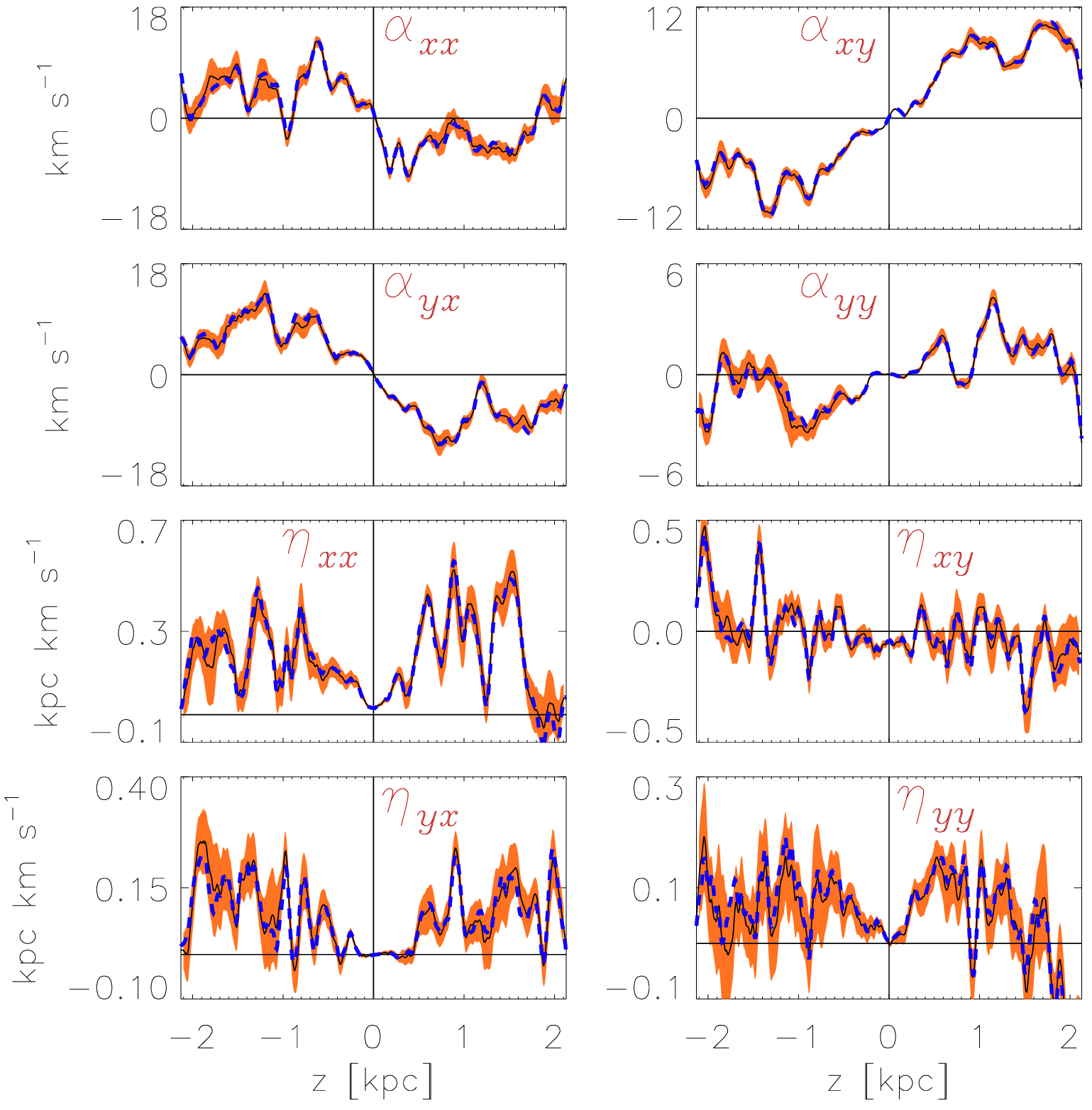}
    \caption{Same as \fref{fig:coeff} but with $m=1$,
    i.e. three points in the local neighbourhood. The 
    dashed-blue line indicate the same coefficients 
    determined with our local version of SVD method 
    discussed in \cite{BK_svd_2020}.}
    \label{fig:coeff_non_local_m_1}
\end{subfigure}
\end{figure*}

Data from the same model was also analysed with the
test-field method previously and dynamo coefficients 
were obtained. Test-field which extends over the 
full $z$-extent of the simulation box (wavenumber 
$k=1$) was used in this analysis. The $z$ profiles 
of the coefficients thus obtained are presented in
\fref{fig:test-field} for reference. To determine 
these, we have divided the kinematic phase (up to
$\sim1$\Gyr) in nine independent time sections, and 
estimated the coefficients for each of these 
sections. Shown in \fref{fig:test-field} with 
black-solid lines are averages of these 
realizations, while the orange shaded regions 
correspond to the square root of the ratio 
of variance in these nine realizations and the 
number of realizations, determined the same way 
as in \fref{fig:coeff} and also in 
\cite{recepies}. Its comparison with the 
results from the local SVD method are discussed in 
\cite{BK_svd_2020}. 

As mentioned in
\sref{sec:results_coefficients} our non-local SVD
calculations with a narrowest possible width of 
local neighborhood (with $m=1$, total 3 points) 
yield the same coefficients as from our local 
calculations. In \fref{fig:coeff_non_local_m_1}
we plot these coefficients along with 
corresponding uncertainties in the determination.
It appears that the coefficients $\alpha_{yy}(z)$ 
and $\gamma(z)$ are comparable with the outcomes 
of test-field method. The differences in the 
determination of other coefficients stem from the 
fact that the test-field method probes these 
coefficients at a fixed wavenumber of the test 
magnetic fields themselves, while in SVD the 
coefficients at all the scales spanned by mean 
fields are determined as a combination. Moreover, 
this set of coefficients does not uniquely 
determine the EMF and covariances associated with 
them, which we have discussed in \cite{BK_svd_2020}.

Furthermore in 
\cite{gressel_test_k_dep}, the authors have 
analyzed the scale dependence of dynamo coefficients 
by varying $k$, considering smaller test-field extent. 
They demonstrated that the smaller the scale of the
test-field, the smaller the transport coefficients,
approximately decreasing with $k$ as a Lorenztian. 
We note that the scale of the mean-field in the DNS 
is a few hundred \pc, which corresponds to larger 
$k$ and the test-field results are then consistent 
with our current non-local determination of these 
transport coefficients.


\bsp	
\label{lastpage}
\end{document}